\documentclass[a4paper,12pt]{article}

\usepackage{times}
\usepackage{microtype} 
\DisableLigatures{encoding = *, family = *}

\usepackage[sort&compress,numbers]{natbib}
\usepackage{doi}

\makeatletter
\renewenvironment{thebibliography}[1]
{\subsubsection*{\refname}%
	\@mkboth{\MakeUppercase\refname}{\MakeUppercase\refname}%
	\list{\@biblabel{\@arabic\c@enumiv}}%
	{\settowidth\labelwidth{\@biblabel{#1}}%
		\leftmargin\labelwidth
		\advance\leftmargin\labelsep
		\@openbib@code
		\usecounter{enumiv}%
		\let\p@enumiv\@empty
		\renewcommand\theenumiv{\@arabic\c@enumiv}}%
	\sloppy
	\clubpenalty4000
	\@clubpenalty \clubpenalty
	\widowpenalty4000%
	\sfcode`\.\@m}
{\def\@noitemerr
	{\@latex@warning{Empty `thebibliography' environment}}%
	\endlist}
\makeatother

\usepackage{graphicx}
\usepackage{dcolumn}
\usepackage{amsmath,amsfonts,amssymb,bm}
\usepackage{color}

\newcommand{\DL}{D_{\rm L}}
\newcommand{\DR}{D_{\rm R}}
\newcommand{\dd}{{\rm d}}
\newcommand{\qa}{\bm q_{\rm a}}
\newcommand{\qb}{\bm q_{\rm b}}
\newcommand{\LL}{L_{\rm L}}
\newcommand{\LR}{L_{\rm R}}

\newcommand{\q}{\bm q}
\newcommand{\x}{\bm x}

\newcommand{\om}{\hat{\bm u}}

\newcommand{\reals}{\mathbb R}

\usepackage[utf8]{inputenc}

\usepackage{ragged2e} 

\usepackage{fancyhdr}
\title{Diffusion within pores fully revealed \\ by magnetic resonance\footnote{For earlier preprints on this technique, the reader is referred to \cite{Ozarslan21ARXIVeverything,Ordinola21ARXIVfirstexp,Ozarslan22ARXIVdemystifying}.}}

\author
{Evren \"Ozarslan,$^{1\ast}$ Cem Yolcu,$^{1}$ Alfredo Ordinola,$^{1}$ Deneb Boito,$^{1}$ \\
	Tom Dela Haije,$^{2}$ Mathias H{\o}jgaard Jensen,$^{2}$ Magnus Herberthson$^{3}$ \\
	\\
	\normalsize{$^{1}$Department of Biomedical Engineering, Link\"oping University, Link\"oping, Sweden}\\
	\normalsize{$^{2}$Department of Computer Science, University of Copenhagen, Copenhagen, Denmark}\\
	\normalsize{$^{3}$Department of Mathematics, Link\"oping University, Link\"oping, Sweden}\\
	\normalsize{$^\ast$To whom correspondence should be addressed; E-mail:  evren.ozarslan@liu.se.}
}

\date{}

\begin{document}
\maketitle

\begin{abstract}
		Probing the transport of fluids within confined domains is important in many areas including material science, catalysis, food science, and cell biology. The diffusion propagator fully characterizes the diffusion process, which is highly sensitive to the confining boundaries as well as the structure within enclosed pores. While magnetic resonance has been used extensively to observe various features of the diffusion process, its full characterization  has been elusive. Here, we address this challenge by employing a special sequence of magnetic field gradient pulses for measuring the diffusion propagator, which allows for `listening to the drum' and determining not only the pore's shape but also diffusive dynamics within it.
\end{abstract}


\vspace{30pt}
The diffusion propagator indicates the probability that a particle located at position $\bm x$ moves to $\bm x'$ between two specified times. The diffusion propagator fully describes the diffusive motion in  environments having restricting or semi-permeable walls, spatially varying diffusivity, external forces, etc. Let us consider a time-invariant diffusion scenario in $d$ dimensions within a closed and connected domain $\Omega$ under the dimensionless potential energy field $U(\x)$. The diffusion propagator for a time interval of duration $t$, $p(\bm x' , t | \bm x)$,  is then the solution to the system of equations
\begin{subequations}\label{eq:main}
	\begin{align}
	\nabla \cdot \left( \mathbf D(\bm x') e^{-U(\x')}  \nabla e^{U(\x')} p (\x',t|\x) \right) & = \frac{\partial p(\x',t|\x)}{\partial t} \label{eq:diffusion} \\
	\lim_{t\rightarrow 0} p(\bm x' , t | \bm x) & = \delta (\bm x' - \bm x) \label{eq:initial} \\
	\hat{\bm{n}} \cdot \mathbf D(\x') \, e^{-U(\x')} \, \nabla  e^{U(\x')} p(\x' , t | \bm x) &=0, \quad \x' \in \partial \Omega \ ,
	\label{eq:robin}
	\end{align}
\end{subequations}
where $\nabla$ is a vector of partial derivatives with respect to the components of $\x'$, and $\hat{\bm{n}}$ is the surface normal at $\x'$.
The first of these is the diffusion equation with diffusion tensor $\mathbf D(\x')$. The initial condition is given by Eq. \eqref{eq:initial}, while the last equation is the reflective boundary condition.
In this example, $U(\x)$, $\mathbf D(\x)$ and $\Omega$ are quantities describing the fluid properties or a static picture of the environment all of which give rise to the particular diffusive dynamics, which is captured by the propagator. If the diffusion propagator is available, the diffusion tensor and the potential landscape can be determined, respectively, from its short-time and long-time behaviors, while $\Omega$ is given by its support.
Clearly, the diffusive process is an indirect yet powerful means of recovering the structure of the medium, making it relevant to many disciplines. 

Magnetic resonance has been the method of choice for many characterization studies due to its noninvasive nature and exquisite sensitivity to diffusion, which has been realized since its early days \cite{Hahn50,Carr54}. In a typical MR experiment, the specimen is subjected to a magnetic field $B_z$ whose direction defines the $z$-axis by convention. The magnetic moments of the spin-bearing particles exhibit coherence, synergistically yielding a magnetization vector that develops along the $z$-axis. By applying electromagnetic radiation at a specific frequency, magnetization due to the nuclei of the atoms of interest can be tilted towards the $xy$-plane, upon which it undergoes Larmor precession at an angular frequency given by $\omega = \gamma B_z$, where $\gamma$ is the gyromagnetic ratio, which is specific to the particular atomic nuclei being examined. Such precession leads to changing magnetic flux around it, inducing a potential difference in a nearby antenna, which is referred to as the MR signal. During the course of the MR experiment, different particles acquire different phase shifts $\left( -\int \omega(\x,t) \, \dd t \right)$ due to the differences in the local magnetic field and experimental manipulations of $B_z$. 

\begin{figure}[t!]
	\begin{center}
		\includegraphics[clip, trim=0cm 0cm 0cm 0cm,width=.67\textwidth]{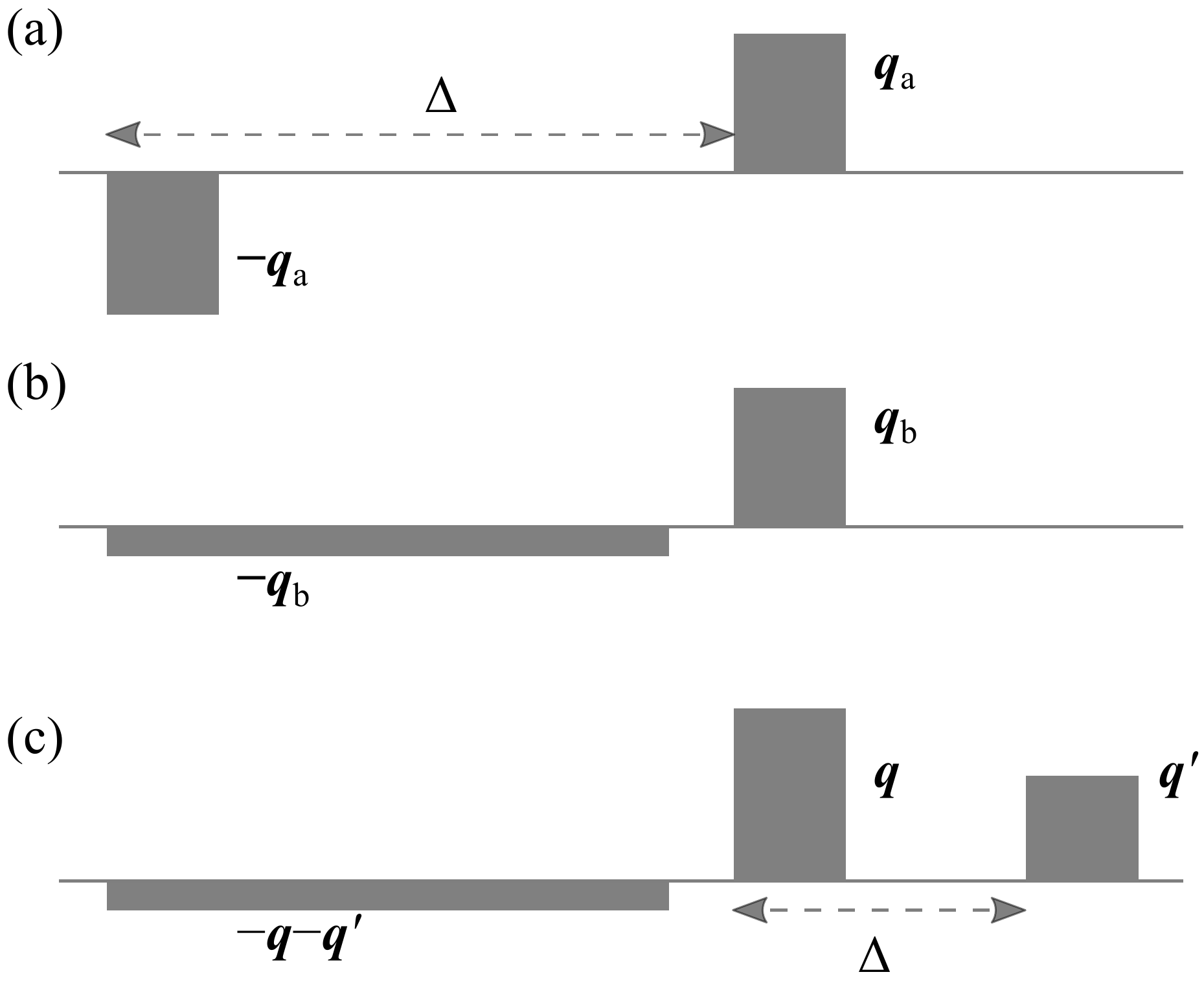}
		\caption{The diffusion encoding pulse sequences considered. (a) Stejskal-Tanner sequence \cite{StejskalTanner65} allows the measurement of the ensemble average propagator. (b) The gradient waveform introduced by Laun et al. \cite{Laun11prl} enables measurement of the long diffusion time limit of the propagator. (c) The pulse sequence introduced here makes it possible to map the diffusion propagator.}
		\label{fig:pulse}
	\end{center}
\end{figure}

One such manipulation introduced by Stejskal and Tanner in 1965 involves incorporating pulsed magnetic field gradients ($\nabla B_z$) into MR acquisitions for performing diffusion measurements in a controllable way \cite{StejskalTanner65}; gradient pulses have also been the building blocks of MR imaging \cite{Lauterbur73}. Stejskal and Tanner's experiment (see Figure \ref{fig:pulse}a) featuring two gradient pulses of equal duration is still the most widely employed diffusion encoding method. Here, $\qa$ denotes the integral of the gradient vector over its duration, multiplied by  $\gamma$. A spin bearing particle, whose average positions during the application of the first and second pulses denoted by $\bm x$ and $\bm x'$, suffers phase shifts of  $\qa \cdot \bm x$ and $-\qa \cdot \bm x'$, respectively, due to the Larmor precession frequency being proportional to the magnetic field. Consequently, the MR signal intensity (divided by the intensity with $\qa = 0$) is given by 
\begin{align} 
E^{({\rm a})}_\Delta(\qa)=\int_\Omega \dd \bm x \, \rho(\bm x) \int_\Omega \dd \bm x' \, p(\bm x', \Delta | \bm x) \, e^{-i \qa\cdot(\bm x' - \bm x )} \ ,
\end{align}
where $\rho(\bm x)$ is the initial spin density and for simplicity, we assumed short pulses that encode the instantaneous positions of the particles. Conventional experiments for measuring self-diffusion start at the steady state, i.e., with $\rho(\bm x')=\lim_{t\rightarrow\infty} p(\bm x', t | \bm x)$ in the absence of sources and relaxation sinks. 
The signal is just the Fourier transform of the ensemble averaged propagator (EAP) defined by
\begin{align}\label{eq:EAP}
\bar P_\Delta(\bm x_\mathrm{net}) = \int_\Omega \dd \bm x \, \rho(\bm x) \, p(\bm x+\bm x_\mathrm{net} , \Delta | \bm x)  \ ,
\end{align}
where $\bm x_\mathrm{net}=\bm x' - \bm x$ is the net displacement vector. Thus, EAP can be computed from the inverse Fourier transform of the signal 
\begin{align} \label{eq:EAP_from_E}
\bar P_\Delta(\bm x_\mathrm{net}) = \frac{1}{(2\pi)^{d}} \int_{{\reals}^d} \dd \qa \, e^{i \qa \cdot \bm x_\mathrm{net}} \,E^{(a)}_\Delta(\qa) \ .
\end{align}

The EAP is a substantially compromised version of the propagator, indicating the likelihood of net displacements averaged for all spins irrespective of where they are within the structure. Despite this limitation, it exhibits very interesting features enabling some understanding of the underlying structure, thus has been widely utilized in characterizing porous media \cite{Karger83,Callaghan91,Mitra92PRL} as well as tissues \cite{Cory90,Wedeen05}.

Recently, Laun et al. introduced another two-pulse experiment, one pulse being long, the other narrow  \cite{Laun11prl} as shown in Fig. \ref{fig:pulse}b. Assuming closed pores and uniform structure within, the particles visit every site within the pore with equal probability during the application of the long pulse. Thus, the positional average of each and every trajectory is very tightly distributed around the pore's center-of-mass. As such, the long pulse has no effect other than diminishing the integral of the waveform, which is a necessary condition for making the signal independent of the pore's position within the specimen. As a result, the signal from all pores add up, generating a detectable signal level even for a specimen comprising small amount of fluid. If the second pulse is short, the sequence  simply introduces a phase shift proportional to each spin's location. The total signal for a connected pore is then given by
\begin{align} \label{eq:E_Laun} 
E^{({\rm b})}(\qb)=\int_{\tilde\Omega} \dd \x \, \tilde\rho(\x) \, e^{-i \qb\cdot \x } \ ,
\end{align}
where $\x$ is the position of the spin with respect to the pore's center-of-mass located at $\x_\mathrm{cm}$ while $\tilde\rho(\x) = \rho(\x+\x_\mathrm{cm})$ and $\tilde\Omega$ indicates the domain translated so that the center of mass of the pore is at the origin.
Thus, the sequence is indeed ``an imaging experiment in disguise,'' \cite{Laun11prl} making it possible to obtain the image of the pore indicator function through an inverse Fourier transform of $E^{({\rm b})}(\qb)$. In more general terms, the obtained quantity is the steady-state distribution of the fluid \cite{Ozarslan17ISMRM}, thus not informative of the diffusion process.

\subsubsection*{Measuring the diffusion propagator}

Here, we consider the sequence in Figure \ref{fig:pulse}c, which combines the key elements of the two sequences discussed above. The long pulse is there so that the integral of the waveform vanishes, and contributions from all pores are independent of their position within the sample. The two subsequent pulses $\bm q$ and $\bm q'$ introduce phase shifts that depend on the particles' positions during their application (in a frame of reference whose origin is at $\x_\mathrm{cm}$---the center of mass of the fluid filling up the pore), denoted by $\bm x$ and $\bm x'$, respectively. When the second and third pulses are short, the signal is given by
\begin{align} 
E^{({\rm c})}_\Delta(\bm q,\bm q')=\int_{\tilde\Omega} \dd \bm x \, \tilde\rho(\bm x) \int_{\tilde\Omega} \dd \bm x' \, \tilde p(\bm x', \Delta | \bm x) \, e^{-i (\bm q\cdot \bm x + \bm q' \cdot \bm x' )} \ ,
\end{align}
where $\tilde p(\bm x', \Delta | \bm x) = p(\bm x' + \x_\mathrm{cm}, \Delta | \bm x + \x_\mathrm{cm})$. 
The propagator, can be obtained via the $2d$-dimensional inverse Fourier transform of the signal
\begin{align}
	W_\Delta(\x,\x') &:= \frac{1}{(2\pi)^{2d}} \int_{\reals^d} \dd \q \, \int_{\reals^d} \dd \q' \, E^{({\rm c})}_\Delta(\q,\q') \, e^{i (\q\cdot\x+\q' \cdot\x')} 
\end{align}
along with an estimate of $\tilde\rho(\mathbf x)$, which is made available by the $d$-dimensional inverse Fourier transform of the subset of the data with $\bm q'=\bm 0$.
In other words, the diffusion propagator is given by
\begin{align} \label{eq:P_c} 
\tilde p(\bm x', \Delta | \bm x) =\frac{\int_{\reals^d} \dd \bm q  \, e^{i \bm q\cdot \bm x } \int_{\reals^d} \dd \bm q' \, e^{i \bm q' \cdot \bm x' } \, E_\Delta^{({\rm c})}(\bm q,\bm q')}{(2\pi)^d \int_{\reals^d} \dd \bm q  \, e^{i \bm q\cdot \bm x }  \, E_\Delta^{({\rm c})}(\bm q,\bm 0)} \ .
\end{align}

\begin{figure}[b!]
	\begin{center}
		\includegraphics[clip, trim=0cm 0cm 0cm 0cm,width=.7\textwidth]{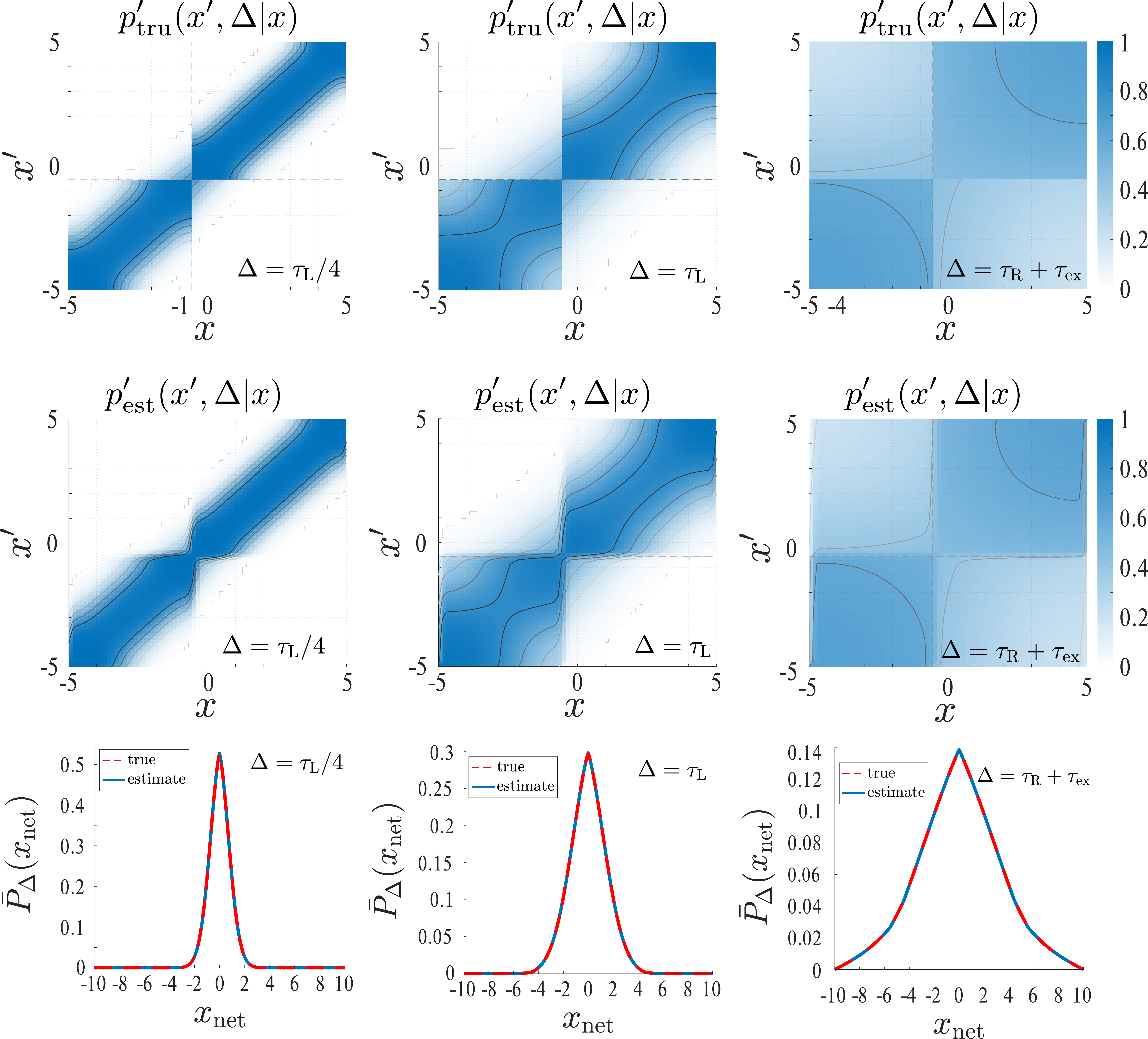}
	\end{center}
	\caption{Representative snapshots of the simulated propagator estimation experiment for exchanging intervals of length $L_{\rm L}$ and $L_{\rm R}$ having diffusivity  $D_{\rm L}$ and $D_{\rm R}$, for the left and right comnpartments, respectively. Shown from top to bottom are: true propagator, estimated propagator, and EAP (true and estimated). The density plots are of $p'(x',\Delta|x) = \tanh \frac {p(x',\Delta|x)}{2/{(L_{\rm L}+L_{\rm R})}}$ for better depiction. The true propagator is computed by its (truncated) spectral decomposition. Membrane position is emphasized by dashed lines. The relaxation time scales $\tau_{\rm L} = \LL^2/\pi^2 \DL$ and $\tau_{\rm R} = \LR^2/\pi^2 \DR$ correspond (roughly) to the process of diffusion within the compartments, and $\tau_{\rm ex} = \sqrt{\DL \DR}/{w}^2$ to the exchange between them.
		\label{fig:validation}}
\end{figure}

\subsubsection*{Structure within the pore}

We demonstrate the estimation of the diffusion propagator of a simulated one-dimensional pore (interval). The pore is partitioned into two exchanging compartments, with diffusion coefficients $\DL$ and $\DR$, separated by a membrane of permeability ${w}$. The walls of the pore are purely reflective. 
The simulations summarized in Figure \ref{fig:validation} illustrates the agreement of the reconstructed propagator (second row) with the true propagators (top row) at three different time intervals.
The associated EAPs are depicted in the bottom row. The presence of a membrane within the pore space is conspicuous  in the estimated propagators while the EAPs are not descriptive.

\subsubsection*{Structural dispersity}

\begin{figure}[h!]
	\begin{center}
		\includegraphics[clip, trim=0cm 0cm 0cm 0cm, width=.44\textwidth]{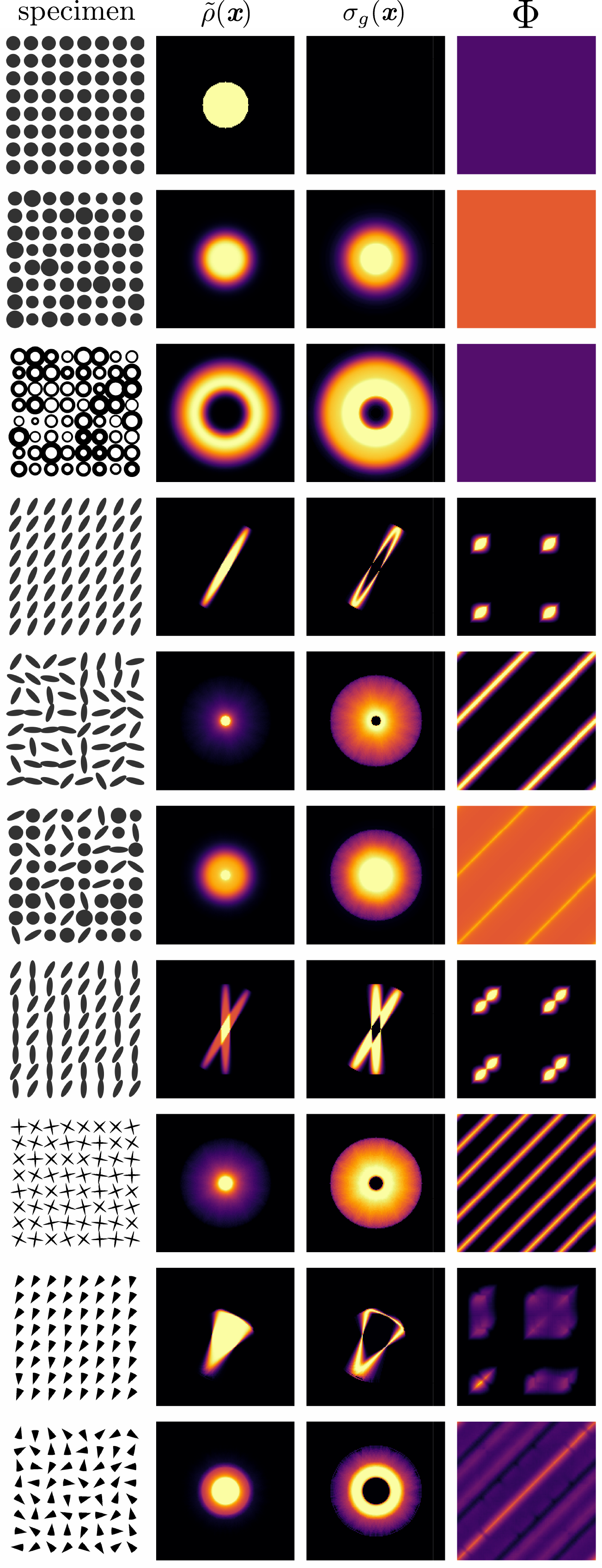}
	\end{center}
	\caption{Maps derived from the long time diffusion measurements for different specimens. The axes of the $\Phi$ maps vary between $-\pi$ and $\pi$. The DI values were estimated to be 0, 0.28, 0.28, $3.8\times 10^{-5}$, $5.0\times 10^{-5}$, 0.41, $3.9\times 10^{-5}$, $4.3\times 10^{-5}$, $1.2\times 10^{-5}$, $1.5 \times 10^{-5}$ (top to bottom). 
	}
	\label{fig:dispersity}
\end{figure}

The propagator-sensitive sequence of Figure \ref{fig:pulse}c can be used to characterize porous media having structural dispersity. To this end, we consider such a specimen having $N$ isolated pores where the $n$th pore has the non-attenuated signal fraction $f_n$. The Fourier transforms of the signals $E_\Delta^{({\rm c})}(\bm q,\bm q')$ and $E_\Delta^{({\rm c})}(\bm q,\bm 0)$  yield 
\begin{subequations}
\begin{align}
W_\Delta(\x,\x')  &= \sum_{n=1,2,3,\ldots}^N f_n \,  \tilde\rho_n (\x) \, \tilde p_n(\x',\Delta|\x) \\
\tilde \rho(\bm x) &=\sum_{n=1,2,3,\ldots}^N f_n \,  \tilde \rho_n (\x) \ , \label{eq:rhobar}
\end{align}
\end{subequations}
which are just weighted averages of the respective quantities for all pores translated so that the pores' centers of mass coincide.  Numerous quantities can be introduced for characterizing the underlying dispersity in the specimen. 
For example, a dimensionless `variance map' can be obtained through the expression 
\begin{align} \label{eq:sigma_g}
\sigma_g(\bm x) := \frac{( W_\infty(\x,\x) - \tilde\rho(\bm x)^2 )^{1/2}}{\rho_\mathrm{max}} \ ,
\end{align}
where $\rho_\mathrm{max}:=\tilde\rho(\x_m)$ is the maximum value of $\rho(\bm x)$.  A dispersity index can be introduced through 
$\mathrm{DI} := \sigma_g^2(\bm x_m) ,$ 
which is equal to $\langle V^{-1} \rangle \langle V \rangle - 1 $ in the absence of external forces when all pores contribute at $\x_m$ and the signal fraction $f_n$ is proportional to the pore volume $V_n$. Here, $\langle \cdot \rangle$ denotes averaging over all pores.
We shall now consider general $\x=r\om$ and $\x'=r'\om'$ where $r=|\x|$, $r'=|\x'|$, and $\om$ and $\om'$ indicate the directions of $\x$ and $\x'$, respectively. One can define the quantity 
\begin{align} \label{eq:Phi}
\Phi(\om,\om') :=\int_0^\infty W_\infty (r\om, r\om') \, r^{2d-1} \, \dd r \ ,
\end{align}
which is constant for a medium composed of isotropic pores. It has peaks at $\om'=\om$ when the pores are anisotropic. If the pore shape has antipodal symmetry, $\om'=-\om$ will exhibit another peak. Similarly, for shapes that exhibit other symmetries (e.g., cross or star-shaped pores) there will be other peaks. In Figure \ref{fig:dispersity}, we illustrate the maps of the quantities in  \eqref{eq:rhobar}-\eqref{eq:Phi} for ten different media with different compositions of two-dimensional pores. 

\subsubsection*{`Hearing the drum'}

\begin{figure}[b!]
	\begin{center}
		\includegraphics[clip, trim=0cm 0cm 0cm 0cm,width=.7\textwidth]{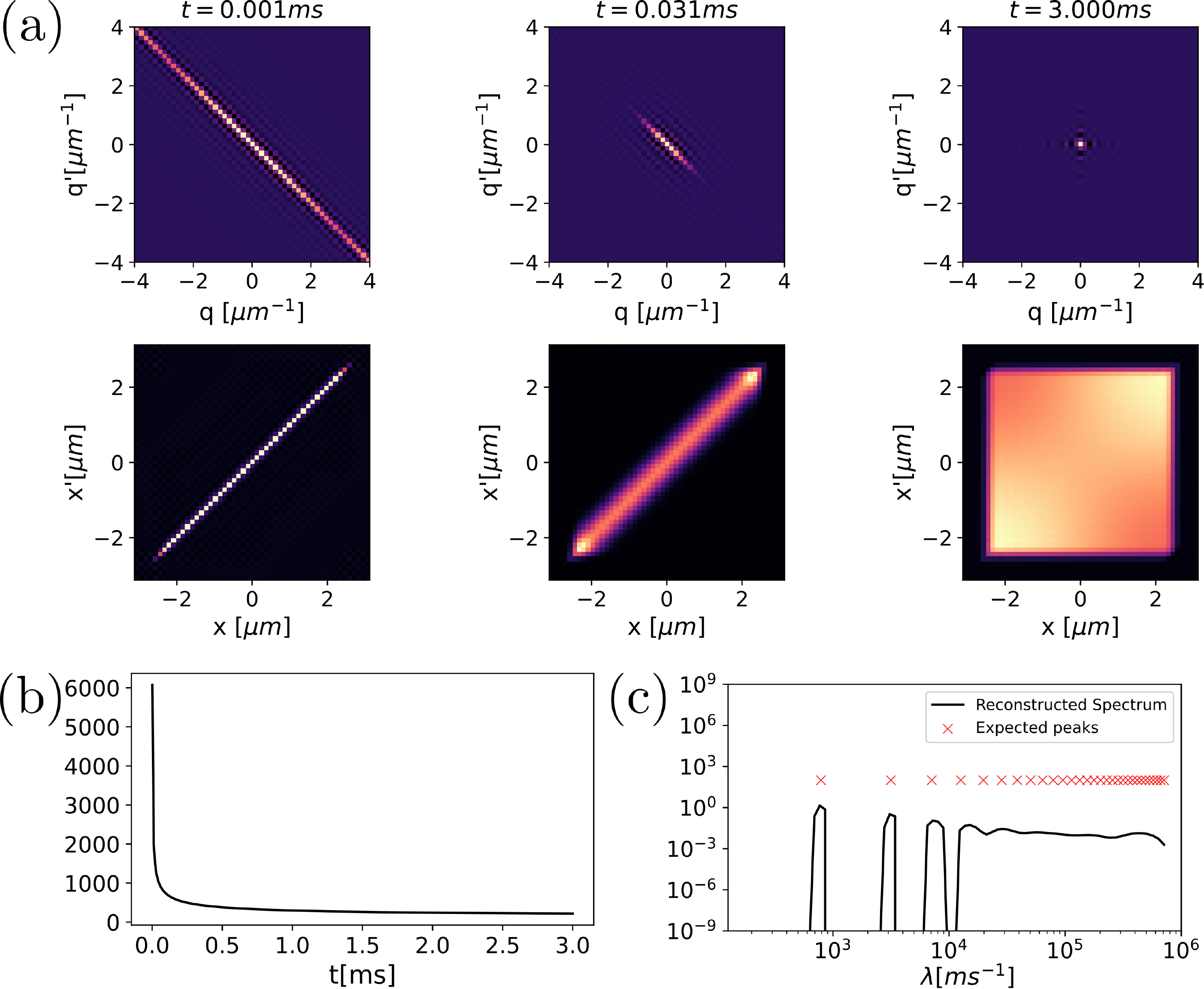}
		\caption{How magnetic resonance ``hears the drum.'' (a) First row: MR signal profiles obtained using the diffusion encoding in Figure \ref{fig:pulse}c. Second row: Propagators obtained via Eq. \eqref{eq:P_c}. (b) Left-hand-side of Eq. \eqref{eq:LaplaceT} plotted against time and (c) its inverse Laplace transform revealing the density of states function. The numerical Laplace inversion \cite{Provencher82} was performed using the package at https://github.com/caizkun/pyilt. }
		\label{fig:hearing}
	\end{center}
\end{figure}

In 1966, Kac posed the now famous question ``Can one hear the shape of a drum?'' \cite{Kac66}, which pertains to recovering the geometry of an enclosing boundary from the eigenspectrum of the Laplacian operator. The pulse sequence of Laun et al. (depicted in Figure \ref{fig:pulse}b) demonstrated that the shape can be recovered from the MR signal.  By enabling the measurement of the diffusion propagator, our gradient waveform illustrated in Figure \ref{fig:pulse}c provides access to the diffusion dynamics within the pore and indeed to the spectrum of the Laplacian. To see this, one can exploit the eigenfunction expansion of the propagator, which leads to the expression
\begin{align} \label{eq:LaplaceT}
\int_\Omega p(\x, t | \x) \, \dd \x = \int_0^\infty g(\lambda) \, e^{-\lambda t} \, \dd \lambda \ . 
\end{align}
Clearly, the density of states, $g(\lambda)$, is accessible from the propagator through an inverse Laplace transform while the propagator is obtained from the signal of the waveform in Figure \ref{fig:pulse}c through Eq. \eqref{eq:P_c}. In Figure \ref{fig:hearing}, we illustrate the recovery of the density of states from simulated signals for the one-dimensional scenario involving diffusion in the direction perpendicular to two parallel plates.

\newpage
\subsubsection*{Concluding remarks}

In conclusion, we introduced a new technique, which facilitates the characterization of the diffusion process within pores in full. This is accomplished by mapping the diffusion propagator through Fourier transforms and can be related to the density of states function making the shape of the drum ``heard'' by magnetic resonance. The technique allows for mapping the structure within closed pores as well as characterizing disperse specimens with unprecedented detail.

\subsubsection*{Acknowledgments}

EÖ thanks Carl-Fredrik Westin for a stimulating conversation, and Nicolas Moutal and Denis Grebenkov for sharing their code on diffusion separated by semi-permeable membranes \cite{Moutal19JSciCom}. 

\small

\end{document}